%% file: main.tex
\documentclass[prl,superscriptaddress,reprint]{revtex4-1}

\usepackage{import}
\import{include/}{preamble-all}

\begin{document}

\title{Relational time in anyonic systems}
\date{\today}
\author{A.~Nikolova}
\email{a.nikolova@uq.edu.au}
\affiliation{School of Mathematics and Physics, University of Queensland, St Lucia, QLD 4072, Australia}
\affiliation{Australian Research Council Centre of Excellence for Engineered Quantum Systems, Australia}
\author{G.~Brennen}
\affiliation{Department of Physics and Astronomy, Macquarie University, Sydney, NSW 2109, Australia}
\affiliation{Australian Research Council Centre of Excellence for Engineered Quantum Systems, Australia}
\author{T.~J.~Osborne}
\affiliation{Institute of Theoretical Physics, University of Hannover, Hannover, Germany}
\author{G.~Milburn}
\author{T.~M.~Stace}
\affiliation{School of Mathematics and Physics, University of Queensland, St Lucia, QLD 4072, Australia}
\affiliation{Australian Research Council Centre of Excellence for Engineered Quantum Systems, Australia}

\begin{abstract}
  \setcitestyle{authoryear,round,aysep={},yysep={,},citesep={;}}
  In a seminal paper~\citep{Page1983} \PW suggest time evolution could be described solely in terms of correlations between systems and clocks, as a means of dealing with the ``problem of time'' stemming from vanishing Hamiltonian dynamics in many theories of quantum gravity. Their approach to \emph{relational time} centres around the existence of a Hamiltonian and the subsequent constraint on physical states. In this paper we present a ``state-centric'' reformulation of the \PW model better suited to theories which intrinsically lack Hamiltonian dynamics, such as \CS theories. We describe relational time by encoding logical ``clock'' qubits into anyons---the topologically protected degrees of freedom in \CS theories. The timing resolution of such anyonic clocks is determined by the universality of the anyonic braid group, with non-universal models naturally exhibiting discrete time. We exemplify this approach using \SU{2}[2] anyons and discuss generalizations to other states and models.
\end{abstract}
\setcitestyle{numeric,square,citesep={,}}

\pacs{}
\keywords{}

\maketitle


\begingroup

In \GR the Hamiltonian is constrained to vanish~\cite{Grav1973}. Canonical quantization preserves this constraint, resulting in the Wheeler--DeWitt equation~\cite{DeWitt1967a}. This equation embodies the ``problem of time'' in canonical quantum gravity: the vanishing of the Hamiltonian on physical states means that all quantum-mechanical operators, including the density matrix describing the state of any system, must be time-independent, in contrast to everyday experience. This apparent paradox has many facets and various approaches attempt to solve some of them (see~\cite{Isham1992,Kuchar2011,Anderson2012} for in depth reviews). One possible solution is that time is relational: that is, it emerges from correlations between subsystems of the Universe, some of which we call ``clocks''.

One of the best known models of this conditional probability interpretation (CPI) is the one proposed by \PW~\cite{Page1983,Wootters1984,PagePOTA} (PaW) and experimentally demonstrated recently~\cite{Moreva2014}. The PaW Universe is formulated in terms of qubits represented as spins, which implicitly carry internal Hamiltonian dynamics. To conform to the Hamiltonian constraint, the state of the Universe is an energy eigenstate, which factors into ``system'' and ``clock'' subspaces.  Then, the ``system'' dynamics emerge with respect to correlations with the ``clock'' subsystem.

In this paper we reformulate the PaW model in a Universe in which there are no implicit Hamiltonian dynamics. Instead, qubits emerge from anyonic degrees of freedom labelling charge sectors of two-dimensional \CS theories that have a vanishing Hamiltonian~\cite{PachosITQC}.

\endignore%
\endgroup


\begingroup

First we give a brief overview of the PaW model. \PW divide the Hilbert space into a ``clock'' part and a ``system'' part, with total Hamiltonian \(\Hop = \Hop\onc + \Hop\ons\), where \(\Hop\oncs\) are Hamiltonians for the clock and system parts, respectively. Following \PW~\cite{Page1983,Wootters1984}, we assume the ``Universe'' is in a pure state, \(\ket[\sysCS]{\Psi_0}\), stationary under a unitary evolution \(\Uop{t} = \exp{-\ii\Hop t}\), \(t\) being the unobservable coordinate time. A reference state, \(\ket[\sysC]{\Tval_0}\), which is not an eigenstate of \(\Hop\onc\), is defined to be the ``zero'' tick of the clock (\ie ``noon'')~\cite{Milburn2005,Boette2016}. Subsequent clock states \(\ket[\sysC]{\Tval}\), are then generated by \(\Hop\onc\),
\begin{align}
  \label{eq:clock-time}
  \ket[\sysC]{\Tval} \coloneqq \ee^{-\ii \Hop\onc \parens{\Tval-\Tval_0}} \ket[\sysC]{\Tval_0}\,,
\end{align}
where \(\Tval\) signifies the ``clock time''. We note that the clock time \(\Tval\) is not associated with any particular value of the coordinate time \(t\); instead it is a possible outcome for a measurement on the clock.

The state of the system at clock time \(\Tval\) is defined by conditioning \(\ket[\sysCS]{\Psi_0}\) on the measured clock state \(\ket[\sysC]{\Tval}\). PaW showed that this conditional state of the system is consistent with Schr\"odinger evolution of the system under \(\Hop\ons\) for a time \(\Tval-\Tval_0\), \ie
\begin{align}
  \label{eq:paw-proj}
  \ket[\sysS]{\psi\of\Tval}	& \coloneqq \frac{ \braket[\sysC]{\Tval}[\sysCS]{\Psi_0} }{ \tr{ \braket[\sysC]{\Tval}[\sysCS]{\Psi_0} \braket{\Psi_0}[\sysC]{\Tval} }^{\sfrac{1}{2}} } \\
  \label{eq:paw-evol}		& = \ee^{-\ii \Hop\ons \parens{\Tval-\Tval_0}} \ket[\sysS]{\psi\of{\Tval_0}}\,.
\end{align}
This is a rather remarkable result relying only on the state being globally---but not locally---stationary, and on the lack of clock--system interactions~\cite{Page1983}. We note that global stationarity leads to problems if a clock is conditioned upon more than once~\cite[Ch.~13]{Kuchar2011}; a point we return to at the end.

\endignore%
\endgroup


\begingroup
\NewRotationOperator{z}{\frac{2\pi\parens{\Tval-\Tval_0}}{\nPOVM}}

The PaW approach outlined above is \emph{Hamiltonian-centric}, in that it starts by defining Hamiltonians for the clock and for the system. \PW then require the joint state be an eigenstate of the total Hamiltonian, \(\Hop\). From there, unitary evolution of the system in clock time, \cref{eq:paw-evol}, follows.

The Hamiltonian-centric approach is conceptually unsatisfying for systems in which the problem of time is manifest. These include \CS theories in which the Hamiltonian vanishes \emph{identically}---a consequence of the \CS Lagrangian being linear in time derivatives~\cite{Dunne1998,PachosITQC}.

Instead, it is more natural to adopt a \emph{state-centric} approach in which we define the joint state, \(\ket[\sysCS]{\Psi_0}\), of the clock--system Universe as well as a canonically ordered set of generalized measurement (POVMs) outcomes for the clock.

Apart from a correlated global state, an ordered POVM of clock states is required for a relational description of time. Whereas in the PaW model the ordering is implicit in the clock Hamiltonian, in our state-centric approach we must impose it explicitly. One convenient way to do this is to introduce a clock Hamiltonian, \(\Hop\onc\), that rotates sequentially between the POVM outcomes. We choose one of the measurement outcomes for the clock as its ``initial'' state, from which the clock ``evolves'' in the manner of \cref{eq:clock-time}. We then find a Hamiltonian, \(\Hop\ons\), for the system partition such that the resulting state after the measurement, \cref{eq:paw-proj}, can be obtained from the initial state, \(\ket[\sysS]{\psi\of{\Tval_0}}\), by evolving it in clock time, as in \cref{eq:paw-evol}. This ensures that the state we started with is an an eigenstate of the total Hamiltonian, \(\Hop = \Hop\onc + \Hop\ons\). We emphasise that in this state-centric formulation of PaW, \(\Hop\onc\) and \(\Hop\ons\) are derived objects.

To exemplify this construction we consider a clock and a system each consisting of a single qubit, prepared in a maximally entangled Bell state,
\begin{align}
  \label{eq:Bell}
  \ket[\sysCS]{\Psi_0} & \coloneqq \half* \parens{\ket{-+} - \ket{+-}}\subsc{\sysCS}\,,
\end{align}
where \(\ket{\pm}\) are the eigenstates of Pauli \(\Xop\).

In line with the spin-\(j\) example in the original PaW paper~\cite{Page1983} we restrict ourselves to clock states on the Bloch sphere's equator (\(\Xax\)--\(\Yax\) plane), and choose \(\ket[\sysC]{\Tval_0} = \ket[\sysC]{+}\). Subsequent clock states are defined by rotations  around the \(\Zax\)~axis,
\begin{align}
  \label{eq:bloch-clock-evol}
  \ket[\sysC]{\Tval} & \coloneqq \rotz[\sysC]\ket[\sysC]{+}\,,
\end{align}
where \(\rotz{\phi} \coloneqq \exp{\ii\phi\ZZ / 2}\), and \(\nPOVM\) is the number of ``ticks'', or possible outcomes, of the clock. To connect with PaW, we observe that this clock time is generated by \(\Hop\onc = -\pi\Zop\onc/\nPOVM\). The time resolution of the clock is \(\Delta\Tval = 2\pi / \nPOVM\), which can be made arbitrarily fine by increasing  \(\nPOVM\).

Conditioning the global state, \cref{eq:Bell}, on the clock state \(\ket[\sysC]{\Tval}\) gives the state of the system at clock time \(\Tval\):
\begin{align}
  \label{eq:paw-evol-Bell}
  \ket[\sysS]{\psi\of\Tval} & = \rotz[\sysS]\ket[\sysS]{-}\,.
\end{align}
By noting that \(\ket[\sysS]{\psi\of{\Tval_0}} = \ket[\sysS]{-}\), \cref{eq:paw-evol-Bell} exactly corresponds to unitary evolution in clock time, \cref{eq:paw-evol}, generated by an effective system Hamiltonian \(\Hop\ons = -\pi\Zop\ons/\nPOVM\).

The state-centric approach is applicable to \CS theories, in which anyons are the source charges. Physical states in these theories are ones that can be prepared by anyon pair-production from the vacuum, braiding and fusion~\cite{PachosITQC}.

\endignore%
\endgroup


\begingroup

We describe anyonic relational time explicitly in the context of the \SU{2}[2] theory. It deals with three particle species, labelled \(\vac\), \(\sigma\), \(\psi\), where \(\vac\) is the vacuum (spin-0 irreducible representation, or irrep), \(\psi\) is a neutral fermion (spin-1 irrep) and \(\sigma\) is the only non-abelian anyon (spin-\(\shalf\) irrep). Measurement of the total topological charge of two \(\sigma\)'s may have more than one possible outcome, as given by the \emph{fusion rules}:
\begin{align}
  \sigma \times \sigma	\to \vac + \psi
  && \sigma \times \psi	\to \sigma
  && \psi \times \psi	\to \vac\,.
\end{align}
The non-deterministic \(\sigma\times\sigma\) fusion rule is what allows a collection of three or more non-abelian anyons to display nontrivial topological degrees of freedom, even when the underlying manifold is contractible~\cite{Kitaev2006,Brennen2009}. These topological degrees of freedom can be used to define qubits, thus enabling clocks in the anyonic PaW universe.

Consider three \(\sigma\) anyons and the associated \emph{fusion Hilbert space}~\footnote{Strictly speaking, due to superselection rules, the Hilbert space is defined by the anyons and their total charge, whatever it may be. The \SU{2}[2] fusion rules constrain the total charge of three \(\sigma\)'s to be \(\sigma\), so we speak of three, rather than four, \(\sigma\)'s as comprising our qubit. In general, for \(n\) anyons in any \SU{2}[k] model, each local subsystem needs to be post-selected on a particular outcome of measurement on the total charge of its anyons such that there are still degrees of freedom associated with intermediate fusion outcomes.}. The order in which we choose to fuse them consecutively defines a basis for this Hilbert space. A given state specifies all intermediate outcomes for that fusion order, and is commonly represented as a labelled tree. We define two possible bases, the ``z'' and ``x'' bases, for fusing three \(\sigma\)'s as \(\cbraces{\ket{\vac\onz},\ket{\psi\onz}}\) and \(\cbraces{\ket{\vac\onx},\ket{\psi\onx}}\), where
\begingroup%
\sibldist=8mm\relax%
\btreeset{%
  defaults,%
  /tikz/btreedefaults,%
  separate,%
  sibling distance scales=false,%
  level distance scales=false,%
}%
\WrapAnyonTree%
\begin{align}
  \label{eq:z-and-x}
  \ket{a\onz} & \coloneqq \ZA{}{\sigma}{a}{\sigma}\,,
  & \ket{a\onx} & \coloneqq \XA{}{\sigma}{a}{\sigma}\,,
\end{align}
\endgroup%
with \(a\sigmain\). We can encode a single qubit in this collective degree of freedom by identifying the ``z'' basis with the computational basis, \(\ket{0} = \ket{\vac\onz}\), \(\ket{1} = \ket{\psi\onz}\). We also define \(\ket{+} = \ket{\vac\onx}\), \(\ket{-} = \ket{\psi\onx}\), so that \(\ket{\pm} = \parens{\ket{0}\pm\ket{1}} / \sqrt2\).

The transformation between the ``x'' and ``z'' basis is given by \(\Fop\), whose elements are determined by the fusion rules:
\begin{align}
  \label{eq:F} 
  \Fop =
    \half*\; \kbordermatrix{
  ~	& \vac	& \psi \\
  \vac	& 1	& 1 \\
  \psi	& 1	& -1  }\,.
\end{align}

Exchanging two \(\sigma\)'s is trivial if their total charge is \(\vac\), and introduces a \(\shalfpi\) phase if their charge is \(\psi\). This is encoded the exchange matrix
\begin{align}
  \label{eq:R}
  \Rop{i,j} =
    \kbordermatrix{
  ~	& \vac	& \psi \\
  \vac	& 1     & 0 \\
  \psi	& 0     & \ii }\,,
\end{align}
given in a basis where the \(i\supth\) and \(j\supth\) \(\sigma\) share a fusion channel~\footnote{Note that the \SU{2}[2] model with the same \(\Fop\) matrix but with \(\Rop \to \ii\Rop^{\dag}\) gives the Ising anyon model.}. The ``y'' basis, \(\cbraces{\ket{\pm\ii}}\), can be defined in terms of the ``z'' basis and braids on anyons 2 and 3 as
\begin{align}
  \ket{+\ii} & \coloneqq \ee^{+\sfrac{\ii\pi}{4}} \Bop{2,3} \ket{1}\,, &
  \ket{-\ii} & \coloneqq \ee^{-\sfrac{\ii\pi}{4}} \Bop{2,3} \ket{0}\,,
\end{align}
where \(\Bop{2,3}\) is given by \( \Bop{2,3} = \Fop* \Rop{2,3} \Fop \).

\begin{figure}
  \centering
  \includegraphics[page=2]{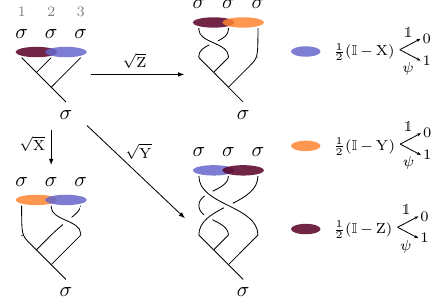}
  \caption{Three \(\sigma\) \SU{2}[2] anyons with total charge \(\sigma\) encode a logical qubit. Pauli measurements, \(\Xop\), \(\Yop\) or \(\Zop\), on the qubit are implemented by fusing a pair of anyons (\ie measuring their total charge), indicated by the coloured ellipses. Fusion of two \(\sigma\)'s yields \(\vac\) or \(\psi\), corresponding to a projective measurement in one of the  Pauli bases \(\cbraces{\Xop,\Yop,\Zop}\), depending on which pair is fused. Braiding among the three anyons effects \(\shalfpi\)-rotations, \(\sqrtXop\), \(\sqrtYop\), \(\sqrtZop\), around the three axes. \(\sqrtXop\), \(\sqrtYop\), \(\sqrtZop\) are equivalent to swapping anyons (2,3), (1,3) and (1,2) respectively.}
  \label{fig:Pauli}
\end{figure}

Qubit measurement is effected by pair-wise anyon fusion (\ie detecting the total charge of a pair, yielding \(\vac\) or \(\psi\)), indicated by coloured ellipses in \cref{fig:Pauli} (top left). The three possible ways to fuse pairs of the \(\sigma\) anyons correspond to measurements in the three Pauli bases \(\Xop\), \(\Yop\), or \(\Zop\).

The braid group of three \(\sigma\) anyons is generated by \(\Rop{1,2}\) and \(\Bop{2,3}\), so it follows that the braid group of \(\sigma\)'s in the \SU{2}[2] model is isomorphic to the one-qubit Clifford group~\cite{Ahlbrecht2009}. Because the Clifford group (braiding), normalises Pauli measurements (fusion), braiding in this model does not give access to additional choices of measurement basis. Thus, projective measurement outcomes on a single, anyonic \SU{2}[2] qubit are restricted to one of the six states, \(\ket{0},\ket{1},\ket{\pm}\) or \(\ket{\pm\ii}\), of which only four are on the Bloch equator. Below, we discuss how this generalises to POVM measurements.

\endignore%
\endgroup


\begingroup

To define relational time in this anyonic Universe, we require
\begin{enumerate*}[label={\roman*)}]
\item at least two subsystems in the Hilbert space,
\item entanglement between the subsystems and
\item a POVM on the clock.
\end{enumerate*}

A minimal anyonic model with two two-dimensional subspaces consists of six \(\sigma\) particles with total charge~\(\vac\)~\cite{Nayak1996}. We define the computational basis as
\begingroup%
\sibldist=12mm\relax%
\begin{align}%
  \begin{tikzpicture}[baseline={(psi.center)}]
    \ZAZBket[atree=psi,scale=1]{\vac}{\sigma}[\sigma]{a}[a']{\sigma,\sigma,\sigma\;,\;\sigma,\sigma,\sigma}
    \node [Left=psi] {\(\ket[\sysCS]{a\onz,a'\onz} \coloneqq {}\)};
    \pic ["{clock,:system,}",node distance=5pt and 5pt] {partition=btreeINNER -0.1 1.3};
  \end{tikzpicture}\,,
  && a,a'\sigmain
\end{align}
\endgroup%

Entanglement requires braiding between the two subsystems. A maximally entangled state is produced when pairs of anyons created from the vacuum are shared between the two subsystems~\cite{Campbell2014} as represented by the following tree:
\begingroup%
\setbox\tmpbox\hbox{\(\sigma\)}%
\tmpdimB=\dimexpr10pt+\labeldist+\dp\tmpbox+\ht\tmpbox\relax%
\sibldist=9.5mm\relax%
\begin{gather}
  \nonumber
  \begin{tikzpicture}[baseline=(current bounding box.center)]
    \ZAZBket[atree=BT]{\vac}{\sigma}[\sigma]{\vac}[\vac]{\sigma, , , , ,\sigma}
    \pic [node distance=20pt and 1pt,"{clock,:system,}"] {partition=btreeINNER -0.1 1.7};
    \settoxdiff{\tmpdimA}{btree-l-l-r}{btree-l-r-r}
    \braid[default=\tmpdimA,omit=2] (braid) at (btree-l-r-r) s_1;
    \node[label above,math] at (braid-1-e) {\sigma};
    \draw [straight braid behind=2];
    \braid[default=\tmpdimA,omit=1] (braid) at (btree-r-r-l) s_1;
    \node[label above,math] at (braid-2-e) {\sigma};
    \draw [straight braid in front=1];
    \settoxdiff{\tmpdimA}{btree-l-r-r}{btree-r-l-l}
    \braid[default=\tmpdimA,omit={1,2}] (braid) at (braid-1-e) s_1;
    \node[label above,math] at (braid-1-e) {\sigma\;};
    \node[label above,math] at (braid-2-e) {\;\sigma};
    \draw [straight braid behind=2,straight braid in front=1];
    \setbox\tmpbox\hbox{\({}=\Rop{3,4}\Bop{4,5}\Bop{2,3}{}\)}
    \node[Right=BT] {\unhcopy\tmpbox};
    \ZAZBket[separate=false,Right=BT \wd\tmpbox]{\vac}{\sigma}[\sigma]{\vac}[\vac]{\sigma,\sigma,\sigma\;,\;\sigma,\sigma,\sigma}%
    \pic [node distance=20pt and 1pt,"{clock,:system,}"] {partition=btreeINNER -0.1 1.7};
  \end{tikzpicture}\\
  \label{eq:Bell-anyons}\equiv \half* \parens{ \ket{+,0} + \ket{-,1} }\subsc{\sysCS} \coloneqq \ket[\sysCS]{\Psi_0}\,.
\end{gather}
\endgroup%

\endignore%
\endgroup


\begingroup

A POVM on the clock can be built by coupling the clock to \(\numa\) ancilla as shown in \cref{fig:circuit}. We initialise the system (\(\sysS\)) and clock (\(\sysC\)) qubits in a Bell state, \cref{eq:Bell-anyons}, and introduce \(\numa\) ancillary qubits. A unitary, \(\Uop\), together with projective measurements on the clock and ancilla, yields a POVM on the clock qubit with \(\nPOVM \le 2^{\numa+1}\) outcomes. In a computationally universal model, for which any unitary \(\Uop\) is physically accessible, the inequality can be saturated, so that the timing resolution of the clock, \(\Delta\Tval = 2\pi / 2^{\numa+1}\), can be made arbitrarily fine by increasing~\(\numa\).

\begin{figure}
  \centering
  \includegraphics[page=2]{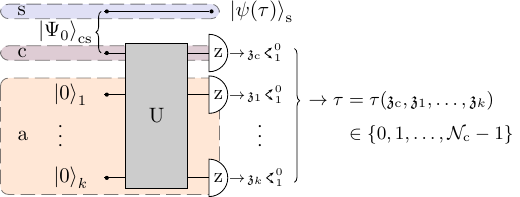}
  \def\bitval{\mathfrak{z}}
  \caption{System qubit, \(\sysS\), and clock qubit, \(\sysC\), are prepared in a Bell state, \(\ket[\sysCS]{\Psi_0}\). To implement a POVM on the clock, the clock is coupled to a collection of \(\numa\) ancilla via a unitary gate, \(\Uop\). Depending on the universality class of the model, \(\Uop\) yields a POVM on the clock qubit with \(\nPOVM \le 2^{\numa+1}\) possible outcomes, which are in direct correspondence with the set of \(\Zop\)-measurement outcomes, \(\cbraces{\bitval_{\sysC},\bitval_1,\ldots,\bitval_{\numa}}\), on the clock and ancilla qubits. An ordering of those outcomes gives the clock time \(\Tval\Tvalin\).}
  \label{fig:circuit}
\end{figure}

\endignore%
\endgroup


\begingroup

The \SU{2}[2] braid group however is isomorphic to the Clifford group, which is \emph{not} universal. In this case, the set of unitary gates generated by the braid group is finite. The maximum number of POVM outcomes, \(\nPOVM\), on the clock is thus bounded: \(\nPOVM \le \nPOVMconst\) for some \(\nPOVMconst\) which depends on the braid group. Time in such a Universe is a discrete quantity, indivisible into intervals smaller than \(2\pi / \nPOVMconst\), regardless of the number of the ancilla used to effect the clock POVM.

The construction here extends to other non-ableian anyonic models. The Universe is modelled as a collection of \(N\) anyons with trivial total charge. We isolate a subset of the fusion Hilbert space
, \(\mathcal{H}\), having \(\numq < N\) degrees of freedom. These degrees of freedom are to be interpreted as qudits, with \(d\) depending on the number of possible fusion outcomes. \(\mathcal{H}\) is split into two non-interacting subsystems---the ``clock'' and the ``system''---such that \(\numq\onc\) ``qudits'' go to the clock while the remaining \(\numq\ons\), to the system. We do this in a way that results in an entangled state of the two subsystems. Clock time is given by an ordered set of POVM outcomes, where the POVM is implemented using \(k\) ancillary qudits. In non-universal models, the temporal resolution, \(\Delta\Tval\), is determined by the braid group and the number of clock qudits, but in all models the resolution is lower bounded by the number of ancilla: \(\Delta\Tval \ge 2\pi / d^{k+1}\).

The connection between the computational universality class of the clock system and the discreteness of relational time is the key result of this Letter. For example the braid group of \SU{2}[4] is not computationally universal, so an \SU{2}[4] Universe would also exhibit discrete relational time (though we note that \SU{2}[4] anyonic models are capable of universal computation under postselection and feedforward~\cite{Cui2015} which is not suitable for defining relational time).

We conjecture that discrete time is generically present in other non-universal theories. This suggests the converse question: are physical theories that exhibit discrete time, including some models of quantum gravity (\eg~\cite{Bojowald2001}), also non-universal for computation? For example, rather than using anyons we could perform our basic protocol using six spin-\(\shalf\) particles with total spin \(\rml{S} = 0\), braiding now being replaced by swap gates, which leave the total spin invariant. This falls into the permutation quantum computation (PQC) model of Marzuoli and Rasetti~\cite{Marzuoli2005} and later Jordan~\cite{Jordan2009}. The PQC model is capable of simulating some processes in the Ponzano--Regge spin foam model of quantum gravity, where \emph{coordinate} time is discretized by performing a Wick rotation to a Euclidean manifold and triangulating that manifold. This is distinct from the \emph{relational} time constructed in this Letter whose discreteness arises from the computational power of the underlying physics of the model Universe.

In addition to discrete time, a non-universal braid group may imply that the Universe it generates does not admit the same level of non-locality as \QM does. A multi-partite state which is nonlocal in standard \QM may admit a local hidden variable theory when the set of allowed measurements is constrained~\cite{Ratanje2011}. Such is the case for the Bell states when only Pauli measurements are allowed, as in the \SU{2}[2] model~\cite{Brennen2009,Ratanje2011,Howard2012,Campbell2014,Clarke2016}. In an \SU{2}[2] Universe at least five \(\sigma\)~pairs (\ie four qubits) shared between two parties are needed to show some non-locality~\cite{Campbell2014}. Thus, while a violation of a CHSH inequality with two qubits implies universality~\cite{Howard2012}\fix{ and vice versa?}, the ability to play some other non-local games is not enough to prove universality and thus the continuity of time.

\endignore%
\endgroup


\begingroup

As a final note, in the PaW model we cannot condition more than once on a clock~\cite[Ch.~13]{Kuchar2011}. One could try to keep relational time flowing after measurement on the clock, by creating a new entangled resource state and teleporting the current system state into a subspace of that resource, using the rest as the new clock~\cite{Brennen2008}. Alternatively, Gambini \etal \cite{Gambini2009} (GPPT) suggest constructing a stationary ``quantum clock'' which is conditioned on a dynamical classical variable, similarly to the way a single system is conditioned on a dynamical clock in the PaW model. The GPPT approach leads to the correct conditional propagators for subsequent measurements on the clock, and might provide a way to recycle anyonic clocks.

\endignore%
\endgroup


\begingroup

We have presented a CPI approach to relational time where qudits are defined in an anyonic fusion space, and where POVMs are generated by braiding and fusion. Our state-centric reformulation of the \PW approach is directly applicable to anyonic models which arise in \CS theories, for which the Hamiltonian vanishes and thus embody the ``problem of time''. We have shown that \SU{2}[k] theories which are non-universal for computation (\ie \(k = 2\) or \(k = 4\)) are only capable of supporting discrete relational time, which may have implications for other models that have discrete, emergent time.

\endignore%
\endgroup


\begin{acknowledgments}
  This work was supported by the Australian Research Council Centre of Excellence for Engineered Quantum Systems (Grant No. CE 110001013). T.~J.~Osborne was supported by the DFG through SFB 1227 (DQ-mat) and the RTG 1991, the ERC grants QFTCMPS and SIQS, and the cluster of excellence EXC201 Quantum Engineering and Space-Time Research.
\end{acknowledgments}


\bibliography{library}

\end{document}

%% file: include/preamble-all.tex
\usepackage{import}

\subimport{.}{preamble-common}
\subimport{.}{preamble-maths}
\subimport{.}{preamble-tikz}
\subimport{.}{preamble-format}

%% file: include/preamble-common.tex
\ifdefined\PRECOMMON\else
\let\PRECOMMON\relax

\usepackage{autocap}		
\usepackage{xparse}		
\usepackage{etoolbox}		

\makeatletter

\newdimen\tmpdimA
\newdimen\tmpdimB
\newbox\tmpbox

\newcount\commentcnt
\commentcnt=0\relax

\def\fix{\textcolor{yellow!70!green!80!black}}

\long\def\beginchange#1\endchange{\begingroup\color{red}#1\endgroup}
\let\endchange\relax

\long\def\beginignore#1\endignore{}
\let\endignore\relax

\let\fix\@gobble

\@ifpackageloaded{tikz}{
  \newcommand*{\comment}[2][]{%
    \begin{tikzpicture}[overlay,remember picture]
      \global\advance\commentcnt by 1\relax
      \coordinate (this-comment) at (0,0 -| current page.east);
      \def\@rotate{-90}
      \def\@anchor{north}
      \pgfextractx\tmpdimA{\pgfpointanchor{current page}{east}}
      \pgfextractx\tmpdimB{\pgfpointanchor{current page}{west}}
      \tmpdimB=-\tmpdimB\relax
      \ifdim\tmpdimA>\tmpdimB\relax
      \coordinate (this-comment) at (0,0 -| current page.west);
      \def\@rotate{90}
      \def\@anchor{north}
      \fi
      \pgfextracty\tmpdimB{\pgfpointanchor{current page}{north}}
      \ifdim\tmpdimA>\tmpdimB\relax
      \coordinate (this-comment) at (0,0 |- current page.north);
      \def\@rotate{0}
      \def\@anchor{north}
      \fi
      \pgfextracty\tmpdimB{\pgfpointanchor{current page}{south}}
      \tmpdimB=-\tmpdimB\relax
      \ifdim\tmpdimA>\tmpdimB
      \coordinate (this-comment) at (0,0 |- current page.south);
      \def\@rotate{0}
      \def\@anchor{south}
      \fi
      \node at (this-comment)
      [anchor=\@anchor,rotate around={\@rotate:(this-comment.center)},
      rectangle callout,callout absolute pointer={(0,0)},callout pointer shorten=0ex,
      draw=red,fill=red,fill opacity=0.2,text opacity=1,text width=\linewidth,#1] {#2};
    \end{tikzpicture}%
  }
}


\def\app@to@hook#1#2{\expandafter\def\expandafter#1\expandafter{#1#2}}
\def\eapp@to@hook#1#2{\edef#1{\unexpanded\expandafter{#1}#2}}
\def\expand@twice{\unexpanded\expandafter\expandafter\expandafter}
\def\apply@macro@to@fixed@numbered@list#1#2#3#4{%
  \begingroup%
  \def\next@el##1##2#3##3\@nil{%
    \eapp@to@hook\tmp@hook{\noexpand#2{%
        \the\numexpr##1\relax}\unexpanded{{##2}}}%
    \ifnum\numexpr##1\relax < \numexpr#1\relax%
      \if\relax\detokenize{##3}\relax%
        \next@el{##1+1}#3\@nil%
      \else\next@el{##1+1}##3\@nil\fi\fi}%
  \def\tmp@hook{}%
  \next@el{1}#4#3\@nil%
  \expandafter\endgroup\tmp@hook%
}
\def\new@macro@factory#1#2{
  \expandafter\def\csname New#1\endcsname{#2\NewDocumentCommand}%
  \expandafter\def\csname Renew#1\endcsname{#2\RenewDocumentCommand}%
  \expandafter\def\csname Declare#1\endcsname{#2\DeclareDocumentCommand}%
}


\NewDocumentCommand{\PW}{ s }{%
  \IfBooleanTF{#1}{%
    Page--Wootters%
  }{%
    Page and Wootters%
  }\ensurespace%
}
\NewDocumentCommand{\SU}{ s m O{} }{%
  SU(#2)%
  \IfBooleanTF{#1}{#3}{\ensuremath{\subsc{#3}}}%
}
\def\CS{Chern--Simons\ensurespace}
\NewAutoCapitalizeMacro{\SR}{special relativity}
\NewAutoCapitalizeMacro{\GR}{general relativity}
\NewAutoCapitalizeMacro{\QM}{quantum mechanics}
\def\etal{et~al.\@\ensurespace}
\def\eg{e.~g.\@\ensurespace}
\def\ie{i.~e.\@\ensurespace}

\makeatother

\fi

%% file: include/preamble-maths.tex
\ifdefined\PREMATHS\else
\let\PREMATHS\relax

\usepackage[all]{math}		
\usepackage{kbordermatrix}	

\usepackage{import}

\subimport{.}{preamble-common}
\makeatletter

\renewcommand{\kbldelim}{(}
\renewcommand{\kbrdelim}{)}


\let\Xop\XX
\let\Yop\YY
\let\Zop\ZZ
\def\sqrtXop{\operatorname{\sqrt{\Xop}}}
\def\sqrtYop{\operatorname{\sqrt{\Yop}}}
\def\sqrtZop{\operatorname{\sqrt{\Zop}}}
\def\Xax{\rml{x}}
\def\Yax{\rml{y}}
\def\Zax{\rml{z}}
\let\ax\rml

\def\sysS{\rml{s}}
\def\sysC{\rml{c}}
\def\sysCS{\rml{cs}}

\DeclareDocumentCommand{\rot}{ s O{} O{} O{} }{
  \IfBooleanTF{#1}{%
    \def\rot@supsc{\dagger}%
  }{%
    \def\rot@supsc{}%
  }%
  \ifstrempty{#2}{}{%
    \expandafter\def\expandafter\rot@supsc\expandafter{\rot@supsc\left( #2 \right)}%
    }%
  \ensuremath{%
    \operatorname{\mathcal{R}}\expandafter\supsc\expandafter{\rot@supsc}\subsc{\ax{#4}}%
    \ifstrempty{#3}{}{\left( #3 \right)}%
  }%
}
\new@macro@factory{RotationOperator}{\@NewRotationOperator}
\DeclareDocumentCommand{\@NewRotationOperator}{ m m G{\alpha} }{%
  \expandafter#1\expandafter{%
    \csname rot#2\endcsname}{ s O{} G{#3} }{%
    \IfBooleanTF{##1}{%
      \rot*[##2][##3][#2]%
    }{%
      \rot[##2][##3][#2]%
    }%
  }%
}
\DeclareDocumentCommand{\Fop}{ s }{
  \operatorname{F}%
  \IfBooleanTF{#1}{\supsc{\dagger}}{}%
}
\DeclareDocumentCommand{\Rop}{ s g }{
  \operatorname{R}%
  \IfBooleanTF{#1}{\supsc{\dagger}}{}%
  \IfNoValueTF{#2}{}{\subsc{#2}}%
}
\DeclareDocumentCommand{\Bop}{ s g }{
  \operatorname{B}%
  \IfBooleanTF{#1}{\supsc{\dagger}}{}%
  \IfNoValueTF{#2}{}{\subsc{#2}}%
}
\DeclareDocumentCommand{\Uop}{ s g }{
  \operatorname{U%
    \IfBooleanTF{#1}{%
      \supsc{\dagger}%
    }{}}%
    \IfNoValueTF{#2}{}{\parens{#2}%
  }%
}
\DeclareMathOperator{\Hop}{H}			

\def\Tval{\tau}					
\def\nPOVM{\mathcal{N}\onc}			
\def\nPOVMconst{\mathcal{M}\onc}		
\def\vac{\mathbbm{1}}				
\def\sigmain{{}\in\cbraces{\vac,\psi}}		
\def\Tvalin{{}\in\cbraces{0,1,\ldots,\nPOVM-1}} 
\def\numa{k}					
\def\numq{n}					

\let\of\parens

\newcommand*{\supth}{\supsc{\text{th}}}
\def\ons{_{\sysS}}
\def\onc{_{\sysC}}
\def\oncs{_{\sysC,\sysS}}
\def\onx{_{\Xax}}

\def\onz{_{\Zax}}

\makeatother

\fi

%% file: include/preamble-tikz.tex
\ifdefined\PRETIKZ\else
\let\PRETIKZ\relax

\usepackage{tikz}		
\usepackage{./binarytree}	
\usepackage{braids}		

\usepackage{import}

\subimport{.}{preamble-common}
\makeatletter

\newdimen\labeldist
\newdimen\sibldist
\newdimen\wrapskip
\labeldist=2.5pt\relax
\sibldist=6.5mm\relax
\wrapskip=0pt\relax

\usetikzlibrary{%
  arrows.meta,%
  calc,%
  fit,%
  intersections,%
  positioning,%
  shapes.callouts,%
  quotes,%
}

\let\AA\sysC
\let\BB\sysS
\def\sysA#1:#2:#3\@nil{#1}
\def\sysB#1:#2:#3\@nil{#2}

\def\settoxdiff#1#2#3{%
  \pgfextractx{#1}{\pgfpointdiff{\pgfpointanchor{#2}{center}}{\pgfpointanchor{#3}{center}}}}
\def\settoydiff#1#2#3{%
  \pgfextracty{#1}{\pgfpointdiff{\pgfpointanchor{#2}{center}}{\pgfpointanchor{#3}{center}}}}

\tikzset{%
  baseline=(current bounding box.center),%
  math/.style={execute at begin node=\(\displaystyle , execute at end node=\)},%
  btreedefaults/.style={%
    external=false,%
    baseline=(current bounding box.center),%
    math labels,%
    label distance=\labeldist,%
    sibling distance=\sibldist,%
    level distance=\sibldist/2,%
  },%
  /pgf/braid/default/.style={border height=0pt,height=#1,width=#1,rotate=180},%
  /pgf/braid/default/.default={\sibldist},%
  /pgf/braid/omit/.style={/pgf/braid/.cd,,style strands={#1}{white}},%
  pics/partition/.style args={#1 #2 #3}{code={
      \path [dashed,draw=black!40] ($ (#1.south) ! #2 ! (#1.north) $) -- ($ (#1.south) ! #3 ! (#1.north) $);
      \node [above left=of #1.north,black!80] {\expandafter\sysA\tikzpictext::\@nil\space \(\AA\)};
      \node [above right=of #1.north,black!80] {\expandafter\sysB\tikzpictext::\@nil\space \(\BB\)};
    }},%
  pics/partition/.default={btree 0 1},%
  /BT/atree/.style={separate=false,local bounding box=#1,%
    execute at end scope={\coordinate (#1-root) at (btree-root);}},%
  label above/.style={inner sep=0pt,minimum size=0pt,above=#1},%
  label below/.style={inner sep=0pt,minimum size=0pt,below=#1},%
  label left/.style={inner sep=0pt,minimum size=0pt,left=#1},%
  label right/.style={inner sep=0pt,minimum size=0pt,right=#1},%
  label above/.default={\labeldist},%
  label below/.default={\labeldist},%
  label left/.default={\labeldist},%
  label right/.default={\labeldist},%
  Above/.style={inner sep=0pt,anchor=south,above=\labeldist of #1.north},%
  Below/.style={inner sep=0pt,anchor=north,below=\labeldist of #1.south},%
  Left/.style={inner sep=0pt,anchor=east,left=0pt of #1.west},%
  Right/.style={inner sep=0pt,anchor=west,right=0pt of #1.east},%
  /BT/Left/.style args={#1 #2}{/tikz/shift={($ (#1-root) - (#1.east |- #1-root) + (#1.west |- #1-root) - (#2,0) $)}},%
  /BT/Right/.style args={#1 #2}{/tikz/shift={($ (#1-root) + (#1.east |- #1-root) - (#1.west |- #1-root) + (#2,0) $)}},%
  /BT/Above/.style args={#1 #2}{/tikz/shift={($ (#1-root) + (0,#2) $)}},%
  /BT/Below/.style args={#1 #2}{/tikz/shift={($ (#1-root) + (0,-#2) $)}},%
  curved braid/.style n args={3}{insert path={%
      (braid-#1-s) .. controls +(0,0.5*#3) and +(0,-0.25*#3) .. (braid-#1-e)}},%
  top curved braid/.style n args={3}{insert path={%
      (braid-#1-s) .. controls +(0.5*#2,0.5*#3) and +(0,-0.25*#3) .. (braid-#1-e)}},%
  bottom curved braid/.style n args={3}{insert path={%
      (braid-#1-s) .. controls +(0,0.25*#3) and +(-0.5*#2,-0.5*#3) .. (braid-#1-e)}},%
  straight braid/.style n args={3}{insert path={%
      (braid-#1-s) -- (braid-#1-e)}},%
  pics/braid/.style n args={3}{code={%
      \begin{scope}[every node/.style={rectangle,inner sep=0pt,outer sep=0pt},shift={(braid-#1-s)}]
        \settoxdiff\pgfutil@tempdima{braid-#1-s}{braid-#1-e}
        \settoydiff\pgfutil@tempdimb{braid-#1-s}{braid-#1-e}
        \node (tl) [fit=(braid-#1-s) ($ (braid-#1-s) ! #2 ! (braid-#1-e) $)] {};
        \node (br) [fit=(braid-#1-e) ($ (braid-#1-e) ! #2 ! (braid-#1-s) $)] {};
        \clip (tl.south west) rectangle (tl.north east) (br.south west) rectangle (br.north east);
        \draw[#3 braid={#1}{\pgfutil@tempdima}{\pgfutil@tempdimb}];
      \end{scope}
    }},%
  curved braid in front/.style={insert path={pic {braid={#1}{1}{curved}}}},%
  curved braid behind/.style={insert path={pic {braid={#1}{0.4}{curved}}}},%
  top curved braid in front/.style={insert path={pic {braid={#1}{1}{top curved}}}},%
  top curved braid behind/.style={insert path={pic {braid={#1}{0.4}{top curved}}}},%
  bottom curved braid in front/.style={insert path={pic {braid={#1}{1}{bottom curved}}}},%
  bottom curved braid behind/.style={insert path={pic {braid={#1}{0.4}{bottom curved}}}},%
  straight braid in front/.style={insert path={pic {braid={#1}{1}{straight}}}},%
  straight braid behind/.style={insert path={pic {braid={#1}{0.4}{straight}}}},%
}


\DeclareDocumentCommand{\WrapAnyonTree}{ s }{%
  \let\BT@draw@treeORIG\BT@draw@tree%
  \def\BT@draw@tree{%
    \begin{scope}[local bounding box=btreeINNER]%
      \BT@draw@treeORIG%
      \pgfmathparse{(\BT@bbox@height + \BT@bbox@depth) / 2}%
      \edef\@@voffset{\pgfmathresult pt}%
      \pgfmathparse{sin(30) * \@@voffset}%
      \edef\@@hoffset{\pgfmathresult pt}%
      \coordinate (NW) at (btree.north west);%
      \coordinate (NE) at (btree.north east);%
      \coordinate (SW) at (btree.south west);%
      \coordinate (SE) at (btree.south east);%
      \IfBooleanTF{#1}{%
        \coordinate (E) at (btree.east);%
        \coordinate (W) at ($ (btree.west) - (\@@hoffset,0) $);%
      }{%
        \coordinate (E) at ($ (btree.east) + (\@@hoffset,0) $);%
        \coordinate (W) at (btree.west);%
      }%
    \end{scope}%
    \draw[use as bounding box] (W) ++(-\wrapskip,0)%
      (NW) -- (W) -- (SW) (SE) -- (E) -- (NE) (E) ++(\wrapskip - 1pt,0);%
  }%
}
\DeclareDocumentCommand{\@anyon@tree@ket}{ m m s }{%
  \begingroup%
  \btreeset{defaults}%
  \IfBooleanTF{#3}{\WrapAnyonTree*}{\WrapAnyonTree}%
  \@anyon@tree{#1}{#2}%
}
\def\@anyon@tree@only{\begingroup\btreeset{defaults}\@anyon@tree}
\def\@anyon@tree@set@final@anyon#1#2{%
  \ifstrempty{#2}{\csletcs{anyon#1}{anyon\the\numexpr#1-1\relax}}{%
    \expandafter\def\csname anyon#1\endcsname{#2}}%
}
\def\@anyon@tree@Z@@path#1#2#3#4{%
  \unexpanded{l:#1!l:#2!l:}%
  \expand@twice{\csname anyon1\endcsname},%
  ll!r:\expand@twice{\csname anyon2\endcsname},%
  lr!r:\expand@twice{\csname anyon3\endcsname}}
\def\@anyon@tree@@Z@path#1#2#3#4{%
  \unexpanded{r:#1!r:#2!r:}%
  \expand@twice{\csname anyon6\endcsname},%
  rr!l:\expand@twice{\csname anyon5\endcsname},%
  rl!l:\expand@twice{\csname anyon4\endcsname}}
\def\@anyon@tree@X@@path#1#2#3#4{%
  \unexpanded{l:#1!r:#2!r:}%
  \expand@twice{\csname anyon3\endcsname},%
  lr!l:\expand@twice{\csname anyon2\endcsname},%
  ll!l:\expand@twice{\csname anyon1\endcsname}}
\def\@anyon@tree@@X@path#1#2#3#4{%
  \unexpanded{r:#1!l:#2!l:}%
  \expand@twice{\csname anyon4\endcsname},%
  rl!r:\expand@twice{\csname anyon5\endcsname},%
  rr!r:\expand@twice{\csname anyon6\endcsname}}
\def\@anyon@tree@Z@Z@path#1#2#3#4{%
  \csname @anyon@tree@Z@@path\endcsname{#1}{#2}\@nil\@nil,
  \csname @anyon@tree@@Z@path\endcsname{#3}{#4}\@nil\@nil}
\def\@anyon@tree@X@X@path#1#2#3#4{%
  \csname @anyon@tree@X@@path\endcsname{#1}{#2}\@nil\@nil,
  \csname @anyon@tree@@X@path\endcsname{#3}{#4}\@nil\@nil}
\def\@anyon@tree@Z@X@path#1#2#3#4{%
  \csname @anyon@tree@Z@@path\endcsname{#1}{#2}\@nil\@nil,
  \csname @anyon@tree@@X@path\endcsname{#3}{#4}\@nil\@nil}
\def\@anyon@tree@X@Z@path#1#2#3#4{%
  \csname @anyon@tree@X@@path\endcsname{#1}{#2}\@nil\@nil,
  \csname @anyon@tree@@Z@path\endcsname{#3}{#4}\@nil\@nil}
\DeclareDocumentCommand{\@anyon@tree}{ m m O{} G{} G{} O{} G{} O{} G{} }{%
  \apply@macro@to@fixed@numbered@list{6}\@anyon@tree@set@final@anyon,{#9}%
  \edef\@path{%
    \if\relax\detokenize{#4}\relax\else\unexpanded{:#4,}\fi
    \csname @anyon@tree@#1@path\endcsname{#5}{#7}{#6}{#8}}%
  \btreeset{%
    separate,%
    /tikz/btreedefaults,%
    #2,#3}%
  \expandafter\BinaryTree\expandafter{\@path}{3}%
  \endgroup%
}
\foreach \lbasis in {X,Z}{%
  \expandafter\xdef\csname \lbasis A\endcsname{%
    \noexpand\@anyon@tree@only{\lbasis @}{%
      sibling distance scales=false,level distance scales=false}}%
  \expandafter\xdef\csname \lbasis B\endcsname{%
    \noexpand\@anyon@tree@only{@\lbasis}{%
      sibling distance scales=false,level distance scales=false}}%
  \expandafter\xdef\csname \lbasis Aket\endcsname{%
    \noexpand\@anyon@tree@ket{\lbasis @}{%
      sibling distance scales=false,level distance scales=false}}%
  \expandafter\xdef\csname \lbasis Bket\endcsname{%
    \noexpand\@anyon@tree@ket{@\lbasis}{%
      sibling distance scales=false,level distance scales=false}}%
  \foreach \rbasis in {X,Z}{%
    \expandafter\xdef\csname \lbasis A\rbasis B\endcsname{%
      \noexpand\@anyon@tree@only{\lbasis @\rbasis}{%
        sibling distance scales,level distance scales}}%
    \expandafter\xdef\csname \lbasis A\rbasis Bket\endcsname{%
      \noexpand\@anyon@tree@ket{\lbasis @\rbasis}{%
        sibling distance scales,level distance scales}}%
  }%
}

\makeatother

\fi

%% file: include/preamble-format.tex
\ifdefined\PREFORMAT\else
\let\PREFORMAT\relax

\usepackage[inline]{enumitem}

\usepackage{xcolor}
\usepackage[T1]{fontenc}
\usepackage[utf8]{inputenc}

\usepackage{hyperref}
\usepackage{cleveref}		

\makeatletter

\binoppenalty=\maxdimen
\relpenalty=\maxdimen
\postdisplaypenalty=\maxdimen

\colorlet{dblue}{blue!50!black}

\hypersetup{			
  bookmarks=true,		
  unicode=false,		
  pdfborder={0 0 0 [0 0]},	
  pdftoolbar=true,		
  pdfmenubar=true,		
  pdffitwindow=false,		
  pdfstartview={FitH},	        
  pagebackref=true		
  pdfnewwindow=true,		
  colorlinks=true,		
  linkcolor=dblue,		
  citecolor=dblue,		
  filecolor=dblue,		
  urlcolor=dblue,		
  breaklinks=true,		
}

\makeatother

\fi